\documentclass[twocolumn,showpacs,preprintnumbers,prb,aps,psfig]{revtex4-1}
\usepackage{amsmath}
\usepackage{graphicx}
\usepackage{dcolumn}
\usepackage{float}

\begin{document}
\title{Ferromagnetic cluster spin-glass behavior in PrRhSn$_{3}$}
\author{V. K. Anand}
\email{vivekkranand@gmail.com, Present address: Ames Laboratory, Department of Physics and Astronomy, Iowa State University, Ames, Iowa 50011, USA.}
\author{D. T. Adroja}
\author{A. D. Hillier}
\affiliation {ISIS Facility, Rutherford Appleton Laboratory, Chilton, Didcot Oxon, OX11 0QX, UK}

\date{\today}

\begin{abstract}
We report the synthesis, structure, and magnetic and transport properties of a new ternary intermetallic compound PrRhSn$_{3}$ which crystallizes in LaRuSn$_{3}$-type cubic structure (space group $Pm\bar{3}n$). At low applied fields the dc magnetic susceptibility exhibits a sharp anomaly below 6~K with an irreversible behavior in zero field cooled (ZFC) and field cooled (FC) susceptibility below 5.5~K. The ac susceptibility exhibits a frequency dependent anomaly revealing a spin-glass behavior with a freezing temperature, $T_{f}$ = 4.3~K. The observation of spin-glass behavior is further supported by a very slow decay of thermo-remnant magnetization (mean relaxation time $\tau$ = 2149~s). However, a small jump at very low field in the isothermal magnetization at 2 K and a weak anomaly in the specific heat near 5.5~K reveal the presence of ferromagnetic clusters. The frequency dependence of the transition temperature $T_{f}$ in the ac susceptibility obeys the Vogel-Fulcher law, $\nu  = \nu _{0}exp[-E_{a}/k_{\rm B}(T_{f}-T_{0})]$ with activation energy $E_{a}/k_{\rm B}$ = 19.1~K. This together with an intermediate value of the parameter $\delta T_{f} = \Delta T_{f}/T_{f}\Delta({\rm log}_{10}\nu)$ = 0.086 provide an evidence for the formation of a cluster-glass state in PrRhSn$_{3}$. Further, we have analyzed the frequency dependence of transition temperature within the framework of critical slowing down, $\tau = \tau_{0} [(T_{f}-T_{\rm SG})/T_{\rm SG})^{-z\nu'}]$ and found the characteristic time constant $\tau_{0}$ = 2.04 $\times$ 10$^{-10}$ s and critical exponent $z\nu'$ = 10.9, which also support a cluster spin-glass behavior in this compound. The magnetic contribution of the specific heat reveals a broad Schottky-type anomaly centered around 10 K and the analysis based on the crystal electric field model indicates a singlet ground state. Further, below $T_f$ the magnetic part of the specific heat exhibits a $T^{3/2}$ temperature dependence. The strong influence of the crystal electric field and a $T^{3/2}$ temperature dependence are also seen in the electrical resistivity which reveals a metallic character and a high magnetoresistance. We also obtain a surprisingly large value of Sommerfeld-Wilson ratio $R_{\rm W}\approx 247$\@.
\end{abstract}


\maketitle

\section{INTRODUCTION}

The recent observation of spin-glass (SG) behavior in the stoichiometric and crystallographically well ordered intermetallic compounds PrAu$_{2}$Si$_{2}$, \cite{Krimmel, Goremychkin} PrRuSi$_{3}$ \cite{Anand} and PrIr$_{2}$B$_{2}$ \cite{Anupam} as well as in URh$_{2}$Ge$_{2}$ (although it possess site disorder on the Rh and Ge sublattices) \cite{Sullow, Sullow2000} have brought new challenges and insights into the mechanism of spin-glass behavior. Our common understanding of mechanism of spin-glass relies on the existence of frustration and disorder, without which spin-glass behavior can not be realized. \cite{Mydosh} It is the crystallographic disorder or a geometrically frustrated lattice that usually frustrates the magnetic moments of a magnetic system; however none of the above Pr-compounds possess any crystallographic disorder, nor do they have geometrically frustrated lattice, as such there is no obvious source of frustration to the magnetic moments. The observation of spin-glass behavior in these crystallographically well ordered compounds is thus very exciting and brings a new perspective to the physics of the spin-glass phenomena.

From a systematic study of the electronic ground state in PrAu$_{2}$(Si$_{1-x}$Ge$_{x}$)$_{2}$ it was concluded that the crystal electric field (CEF) plays an important role in destabilizing the magnetic moments. \cite{Goremychkin} A novel mechanism due to dynamic fluctuations of the crystal field levels has been proposed for the spin-glass behavior in PrAu$_{2}$Si$_{2}$.\cite{Goremychkin} The inelastic neutron scattering study of PrAu$_{2}$Si$_{2}$ revealed a CEF-split singlet ground state and it was found that the exchange coupling is very close to the critical value for the induced moment magnetism. Therefore the induced moment magnetism in PrAu$_{2}$Si$_{2}$ is destabilized by the dynamic fluctuations of crystal-field level, thus resulting in a frustrated magnetic ground state. PrRuSi$_{3}$ also has a CEF-split singlet ground state and the spin-glass behavior in PrRuSi$_{3}$ is also believed to have its origin in dynamic fluctuations of the crystal field levels. \cite{Anand} However, the origin of the spin-glass behavior in PrIr$_{2}$B$_{2}$ is not quite clear at the moment but the presence of a CEF-split singlet ground state is inferred from the specific heat data which is in line with the underlying mechanism of crystal field induced frustration. \cite{Anupam} Here we present another intermetallic compound PrRhSn$_{3}$ which is crystallographically well ordered and reveal a cluster spin-glass transition at 4.3 K.

PrRhSn$_{3}$ belongs to the family of ternary intermetallic compounds RTX$_{3}$ (R = rare earths, T = transition metals, X = Si, Ge, Sn, Al, Ga) which are known to present diverse magnetic and superconducting properties. \cite{Yama, Pikul, Naka, Pecharsky, Krishna, Conti, Muro, Settai, Furukawa, Kimura, Kimura2007, Sugitani, Tada, AnandSSC, AnandPRB, Hillier, AnandJPCM} Recently we started working on the RTX$_{3}$ compounds and investigated the magnetic and transport properties of PrNiGe$_{3}$, \cite{AnandSSC} PrRhGe$_{3}$, \cite{AnandSSC} PrRuSi$_{3}$, \cite{Anand} LaRhSi$_{3}$, \cite{AnandPRB} CeRhGe$_{3}$, \cite{Hillier} and CeRhSn$_{3}$. \cite{AnandJPCM} In our recent investigation of the Kondo lattice compound CeRhSn$_{3}$ we found a complex magnetic ground state in this compound, a possible ferrimagnetically orderd state below 4 K and a transition from the ferri- to a ferro-magnetic order below 1 K.\cite{AnandJPCM}  We also observed a new kind of frequency dependence in the ac susceptibility measurement where the transition temperature was found to decrease with increasing frequency whose origin is not clear at the moment. Continuing our work on RTX$_{3}$ compounds we have investigated the Pr-analog of CeRhSn$_{3}$, PrRhSn$_{3}$ which like CeRhSn$_{3 }$ also forms in LaRuSn$_{3}$-type cubic structure (space group $Pm\bar{3}n$) in which R atoms occupy two different crystallographic sites. \cite{Eisenmann} Here we report the results of our investigations of the magnetic and transport properties of PrRhSn$_{3}$ through the ac and dc magnetic susceptibilities, isothermal magnetization, thermo-remnant magnetization, specific heat, and electrical resistivity studies. The observation of cluster-spin glass transition in this well stoichiometric compound is very exciting and is expected to enrich our understanding of mechanism behind the spin-glass transition which are not induced by the crystallographic disorder or geometrical frustration. Pr$_3$Rh$_4$Sn$_{13}$, which is closely related to PrRhSn$_{3}$ in stoichiometry and structure, is reported to exhibit no phase transition down to 0.2~K.\cite{Oduchi}

\section{Experimental}

A polycrystalline sample of PrRhSn$_{3}$ was prepared by the standard arc melting technique using the high purity elements (Pr:  99.9\%, Rh: 99.99\%, Sn 99.999\%). Pr, Rh and Sn were taken in the stoichiometric 1:1:3 ratio and arc melted on a water cooled copper hearth under the titanium gettered inert argon atmosphere. During the melting process the arc melted button was flipped and re-melted several times which improved the homogeneity and reaction among the constituents. The as-obtained ingot of PrRhSn$_{3}$ was then wrapped in tantalum foil and annealed for a week at 900 $^\circ$C under dynamic vacuum. The crystal structure and the phase purity of the annealed sample was checked by the powder X-ray diffraction (XRD) using the copper K$_{\alpha}$ radiation, and scanning electron microscopy (SEM). The dc magnetization was measured by using a commercial superconducting quantum interference device (SQUID) magnetometer (MPMS, Quantum-Design). The specific heat was measured using the relaxation method in a physical properties measurement system (PPMS, Quantum-Design).  The electrical resistivity was measured by the standard four probe ac technique in the PPMS. The ac susceptibility was also measured in the PPMS.

\section{Results and discussion}

\begin{figure}
\includegraphics[width=8.5cm]{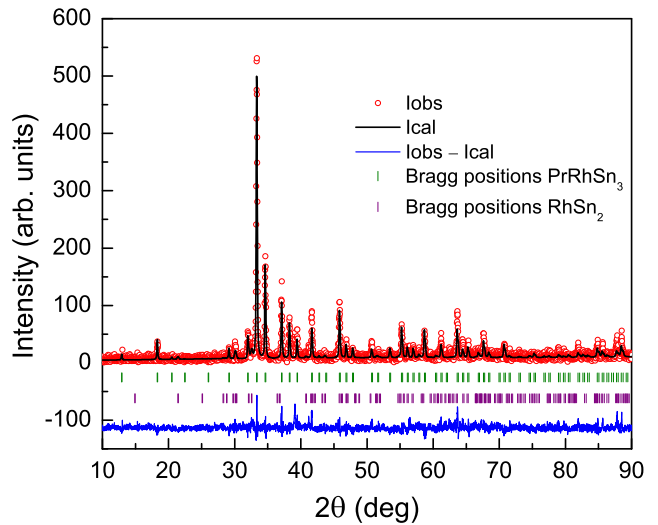}
\caption{\label{fig:fig_XRD} (Colour online) The powder X-ray diffraction pattern of PrRhSn$_{3}$ recorded at room temperature. The solid line through the experimental points is the two-phase Rietveld refinement profile calculated for LaRuSn$_{3}$-type cubic (space group $Pm\bar{3}n$) and CoGe$_2$-type orthorhombic (space group $Cmca$) structural model. The short vertical bars indicate the Bragg peak positions of both the phases. The lowermost curve represents the difference between the experimental and model results.}
\end{figure}

\begin{table}
\caption{\label{tab:table1} Crystallographic and refinement parameters obtained from the structural refinement of powder x-ray diffraction data of PrRhSn$_{3}$. The refinement quality parameter  $\chi^2$ = 1.60.}
\begin{ruledtabular}
\begin{tabular}{lcccc}
\multicolumn{2}{l}{Structure} & \multicolumn{2}{l} {LaRuSn$_{3}$-type cubic}\\
\multicolumn{2}{l}{Space group} & \multicolumn{2}{l} {$Pm\bar{3}n$, No. 223} \\
\multicolumn{2}{l}{Formula units/unit cell}  & \multicolumn{1}{l} {8}\\
\multicolumn{2}{l}{Lattice parameters}\\
 \multicolumn{1}{l}{\hspace{0.8 cm}  $a$ (\AA)}    &    &  \multicolumn{1}{l} {9.6916(8)} \\
 \multicolumn{1}{l}{\hspace{0.8 cm}  $V_{\rm cell}~{\rm (\AA^{3})}$} & &  \multicolumn{1}{l} {910.3(1)}  \\
\multicolumn{2}{l}{Atomic coordinates} \\
 \hspace{0.7cm} Atom & Wyckoff &	 $x$ 	&	$y$	&	$z$	  \\	
 & symbol \\			
	\hspace{0.8 cm}	Pr1 &  2a 	& 0 	& 0          & 0\\
	\hspace{0.8 cm}    Pr2 &  6d 	& 1/4 	&1/2         & 0 \\
	\hspace{0.8 cm}    Rh &   8e 	& 1/4 	& 1/4         & 1/4\\
    \hspace{0.8 cm} 	Sn 	&  24k 	& 0 	& 0.3092(6)   & 0.1532(7)\\
\end{tabular}
\end{ruledtabular}
\end{table}

The X-ray diffraction (XRD) data collected on the powdered polycrystalline sample of PrRhSn$_{3}$ at room temperature were analyzed by Rietveld structural refinement using the software {\tt FullProf}.\cite{Rodriguez} Figure~\ref{fig:fig_XRD} shows the XRD pattern of PrRhSn$_{3}$ together with the Rietveld fit profile. The Rietveld refinement of the XRD data reveals an almost single phase nature of the sample with LaRuSn$_{3}$-type cubic structure (space group $Pm\bar{3}n$, No. 223). The crystallographic parameters obtained from the refinement are listed in Table~\ref{tab:table1}. The Rietveld refinement also revealed a small amount of impurity phase(s). We have identified the impurity phase to be RhSn$_2$ (with CoGe$_2$-type orthorhombic (space group $Cmca$) structure),\cite{Schubert} having a volume fraction of 1.18\% which is equivalent to 2.03\% in molar fraction. We believe that such a small amount of impurity should have no consequence on the results discussed here. The high resolution SEM images also revealed an almost single phase nature of sample.

\begin{figure}
\includegraphics[width=8cm]{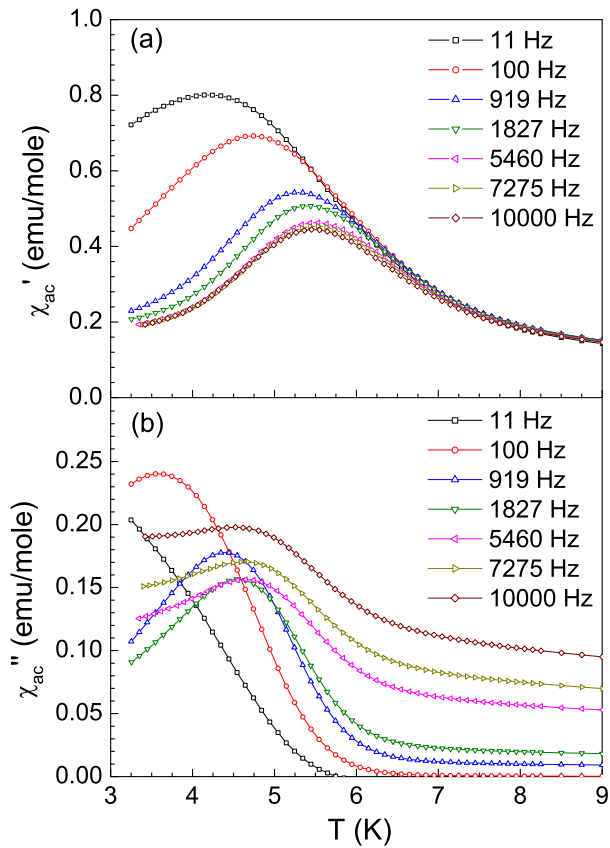}
\caption{\label{fig:fig_Chi_ac} (Colour online) The temperature dependence of the real and imaginary parts of the ac magnetic susceptibility ($\chi_{ac}$) of PrRhSn$_{3}$ measured at different frequencies from 11~Hz to 10~kHz in an applied ac magnetic field of 1.5~mT.}
\end{figure}

Figure~\ref{fig:fig_Chi_ac} shows the temperature dependence of the real ($\chi'_{ac}$) and imaginary ($\chi''_{ac}$) parts of the ac magnetic susceptibility of PrRhSn$_{3}$ measured in an excitation field of 1.5~mT at various frequencies ($\nu$). Both $\chi'_{ac}$ and $\chi''_{ac}$ exhibit pronounced anomalies whose amplitude and peak position depend on the frequency of the applied ac magnetic field. The position of the maxima shifts to higher temperatures as the frequency is increased. As seen from Fig.~\ref{fig:fig_Chi_ac}(a) $\chi'_{ac}$ has a maximum near 4.34 K at low frequency (11 Hz) which shifts to 5.55 K at high frequency (10 kHz). It is seen that the amplitude of maxima in $\chi'_{ac}$ decreases with increasing frequency. However, in the case of $\chi''_{ac}$ the amplitude of maxima initially decreases up to 1827 Hz, above which it starts increasing with further increase in frequency. This effect of frequency on the amplitude of maxima in $\chi''_{ac}$ is unusual. However, the magnitude of $\chi''_{ac}$ at a temperature above the anomaly continuously increases with increasing frequency.

Thus it is seen that the ac susceptibility anomaly strongly depends on the excitation frequency, the position of maxima increases with the increasing frequency which is a characteristic feature of a spin-glass ordering and hence indicates a spin-glass type transition in PrRhSn$_{3}$ with a freezing temperature $T_{f} \sim$ 4.3~K. A criterion which is often used to compare the frequency dependence of $T_{f}$ in different spin-glass systems is to compare the relative shift in freezing temperature per decade of frequency,
\begin{equation}
\delta T_{f} = \frac {\Delta T_{f}}{T_{f} \Delta({\rm log}_{10}\nu)}.
\end{equation}
For PrRhSn$_{3}$ we found $\delta T_{f}$ = 0.086 which is intermediate between those reported for the canonical spin-glass systems (e.g., $\delta T_{f}$ $\sim $ 0.005 for CuMn), and those reported for noninteracting ideal superparamagnetic systems (e.g., $\delta T_{f}$ $\sim $ 0.28 for holmium borate glass a-[Ho$_2$O$_3$(B$_2$O$_3$]). \cite{Mydosh} Though the value of relative frequency shift $\delta T_{f}$ of our compound is considerably larger than that of metallic spin glasses, it is comparable to that of insulating spin glasses.\cite{Mydosh} The value of $\delta T_{f}$ = 0.086 better characterizes our system as the so-called cluster-glass.

\begin{figure}
\includegraphics[width=8.5cm]{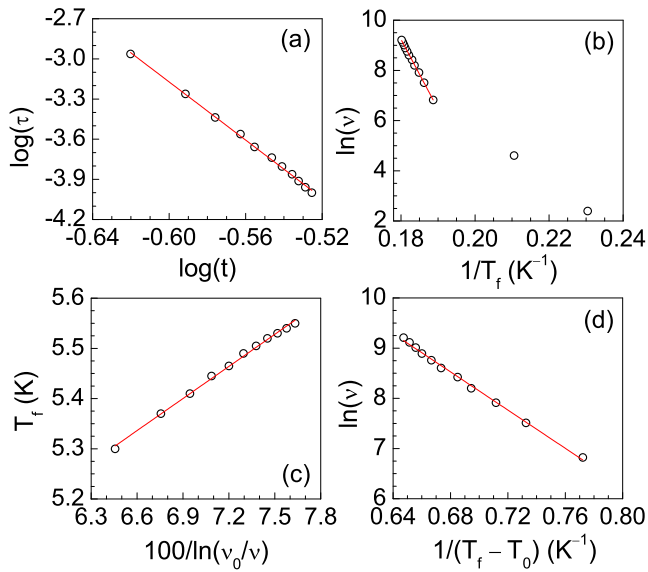}
\caption{\label{fig:fig_Tf} (Colour online) (a) The frequency dependence of freezing temperature plotted as a ${\rm log}(\tau)$ \textit{vs.} ${\rm log}(t)$, where reduced temperature $t = (T_{f}-T_{\rm SG})/T_{\rm SG}$. The solid line represents the fit to the power law divergence. (b) The frequency dependence of freezing temperature plotted as ${\rm ln}(\nu )$ \textit{vs}. $1/T_{f}$. The solid line represents the fit to Arhennius law. (c) The frequency dependence of freezing temperature plotted as $T_{f}$ \textit{vs}. $100/{\rm ln}(\nu_{0}/\nu $). The solid line represents the fit to Vogel-Fulcher law. (d) The frequency dependence of freezing temperature plotted as ${\rm ln}(\nu)$ \textit{vs}. $1/(T_{f} - T_{0})$ together with the fit to Vogel-Fulcher law.}
\end{figure}

Figure 3 shows the frequency dependence of freezing temperature $T_{f}$ obtained from the real part of the ac susceptibility. The frequency dependence of $T_{f}$ follows the conventional power law divergence of critical slowing down, \cite{Mydosh,Hohenberg,Binder}
\begin{equation}
\tau = \tau_{0} \left( \frac {T_{f}-T_{\rm SG}}{T_{\rm SG}} \right)^{-z\nu'},
\label{eq:Power-law}
\end{equation}
where $\tau$  is the relaxation time corresponding to the measured frequency $(\tau = 1/ \nu)$, $\tau_{0}$ is the characteristic relaxation time of single spin-flip, $T_{\rm SG}$ is the spin glass temperature as frequency tends to zero, and $z\nu' $ is the dynamic critical exponent [$\nu' $ is the critical exponent of correlation length, $\xi = (T_f/T_{\rm SG} - 1)^{-\nu'}$ and the dynamical scaling relates $\tau$ to $\xi $ as $\tau \sim \xi^z$]. For a spin-glass system the critical exponent $z\nu' $ is typically found to lie between 4 and 12. It is useful to rewrite Eq.~\ref{eq:Power-law} as
\begin{equation}
{\rm log}(\tau) = {\rm log}(\tau_{0}) -z\nu' ln(t),
\label{eq:Power-law-fit}
\end{equation}
where $t = (T_{f}-T_{\rm SG})/T_{\rm SG}$. The slope and intercept of ${\rm log}(\tau)$ {\it vs}. ${\rm log}(t)$ plot can thus be used to estimate $\tau_{0}$ and $z\nu'$. A log-log plot of inverse frequency ($\tau$) \textit{vs}. reduced temperature ($t$) of PrRhSn$_{3}$ is shown in Fig.~\ref{fig:fig_Tf}(a). The value of $T_{\rm SG}$ was determined by extrapolating the $T_{f}$ \textit{vs}. $\nu $ plot to $\nu $ = 0, which gives $T_{\rm SG}$ = 4.28 K. The solid line in Fig.~\ref{fig:fig_Tf}(a) represents the fit to the power law divergence (Eq.~\ref{eq:Power-law-fit}). The best fit was obtained with $\tau_{0} = 2.04 \times 10^{-10}$~s and $z\nu'  = 10.9(2)$. The value of $z\nu' $ is consistent with spin-glass behaviour in PrRhSn$_{3}$. However, the value of $\tau_{0}$ is large compared to the typical value of 10$^{-12 }$ s for canonical SG systems, which suggests a slow spin dynamics in PrRhSn$_{3}$ likely due to the presence of strongly interacting clusters rather than individual spins.

The presence of interacting clusters is also evident from the departure of frequency dependence of $T_{f}$ from the Arrhenius law,\cite{Mydosh,Binder}
\begin{equation}
 \nu  = \nu_{0} exp\left(-\frac{E_{a}}{k_{\rm B}T_{f}}\right)
\end{equation}
where $k_{\rm B}$ is the Boltzmann constant, $\nu _{0}$ is the characteristic attempt frequency and $E_{a}$ is the average thermal activation energy. According to this law one would expect a linear behavior in a plot of ${\rm ln}(\nu)$ against $1/T_{f}$. However, it can be seen from the ${\rm ln}(\nu)$ {\it vs}. $1/T_{f}$  plot (Fig.~\ref{fig:fig_Tf}(b)) that there is significant deviation from the expected linear behavior at low frequencies, implying that the dynamics is not simply associated with the single-spin flips, rather reflects a cooperative character of the freezing-in process in PrRhSn$_{3}$. We made an attempt to estimate the activation energy $E_{a}$ by fitting the data above 1800 Hz to the Arhenius law (the solid line in Fig.~\ref{fig:fig_Tf}(b)), yielding $E_{a}/k_{\rm B} = 278$~K and $\nu_{0} \sim $ 10$^{26}$~Hz which are totally unphysical. However, the frequency dependence of freezing temperature $T_{f}$ when fitted to the empirical Vogel-Fulcher law, \cite{Mydosh, Vogel, Fulcher, Tholence, Souletie}
\begin{equation}
\nu  = \nu _{0} exp\left(-\frac{E_{a}}{k_{\rm B}(T_{f}-T_{0})}\right)
\label{eq:Vogel-Fulcher}
\end{equation}
with three fitting parameters -- the characteristic attempt frequency $\nu _{0}$, the activation energy $E_{a}$, and the Vogel-Fulcher temperature $T_{0}$ which is often interpreted as a measure of intercluster interaction strength -- gives a reasonable estimate of activation energy. We have tried to estimate the fitting parameters $\nu _{0}$, $E_{a}$, and $T_{0}$ in two different ways, both of which gave consistent values of these parameters. First we fixed the value of attempt frequency to $\nu _{0} = 1/\tau_{0}$, the characteristic time constant $\tau_{0} = 2.04 \times 10^{-10}$~s as determined above and fitted the data. In order to fit the data it is convenient to rewrite Eq.~\ref{eq:Vogel-Fulcher} as
\[
{\rm ln}\left(\frac{\nu_0}{\nu}\right) = \frac{E_{a}}{k_{\rm B}(T_{f}-T_{0})},
\]
which can be rearranged to
\begin{equation}
T_f = \frac{E_{a}/k_{\rm B}}{{\rm ln}(\nu_0/\nu)} + T_0.
\label{eq:Vogel-Fulcher-fit1}
\end{equation}
Thus $E_{a}/k_{\rm B}$ and $T_0$ can be obtained from the slope and intercept of $T_{f}$ \textit{vs}. $1/{\rm ln}(\nu _{0}/\nu)$ plot. A plot of $T_{f }$ \textit{vs}. $100/{\rm ln}(\nu _{0}/\nu )$ together with the fit to Vogel-Fulcher law (Eq.~\ref{eq:Vogel-Fulcher-fit1}) is shown in Fig.~\ref{fig:fig_Tf}(c). The best fit was obtained with $E_{a}/k_{\rm B} = 21.3(4)$~K and $T_{0} = 3.93(3)$~K. In order to make sure that we are not getting wrong fitting parameters as a consequence of fixing the attempt frequency $\nu_{0}$, we determined the value of Vogel-Fulcher temperature $T_{0}$ by following the method suggested by Souletie and Tholence \cite{Souletie} and obtained $T_{0} = 4.01$~K which we used to find $\nu_{0}$ and $E_{a}$. For this we rearrange Eq.~\ref{eq:Vogel-Fulcher} as
\begin{equation}
{\rm ln}(\nu) = {\rm ln}(\nu_{0}) - \frac{E_{a}/k_{\rm B}}{T_{f}-T_{0}}
\label{eq:Vogel-Fulcher-fit2}
\end{equation}
which would then allow us to estimate $E_{a}/k_{\rm B}$ and $\tau_0$ from the slope and intercept of  ${\rm ln}(\nu)$ \textit{vs}. $1/(T_{f} - T_{0})$ plot, respectively. A plot of ${\rm ln}(\nu)$ \textit{vs}. $1/(T_{f} - T_{0})$ is shown in Fig.~\ref{fig:fig_Tf}(d). A linear fit to ${\rm ln} (\nu)$ \textit{vs}. $1/(T_{f} - T_{0})$ plot (the solid line in Fig.~\ref{fig:fig_Tf}(d)) gives characteristic relaxation time, $\tau_{0} = 4.72 \times 10^{-10}$~s and activation energy $E_{a}/k_{\rm B} = 19.1(4)$~K\@. We thus obtain a reasonable estimate of the activation energy, $E_{a} \sim 4.4\,k_{\rm B}T_{f}$ from the peak temperature $T_{f}$ in $\chi'_{ac}$.

Thus we see that the $T_0$ is nonzero in PrRhSn$_{3}$. A nonzero value of $T_0$ arises from the interaction between the spins and indicates the formation of clusters. As such, a nonzero value of $T_0 $ obtained for our compound suggests a cluster spin-glass behavior in PrRhSn$_3$. Further, the fact that $T_{0}$ is very close to $T_{f}$ suggests that the RKKY interaction is relatively strong in our compound. The Tholence criterion \cite{Tholence1984}
\begin{equation}\label{RKKY}
  \delta T_{\rm Th} = \frac{T_f-T_0}{T_f} = 0.076
\end{equation}
for our compound which is obtained using the values $T_f = 4.34$~K (for $\nu = 11$~Hz) and $T_{0}= 4.01$~K\@. This value of $\delta T_{\rm Th}$ is comparable to that of RKKY spin-glass systems (e.g., $\delta T_{\rm Th}=0.07$ for CuMn system).\cite{Tholence1984} Altogether these observations suggest that PrRhSn$_{3}$ falls in the category of RKKY spin-glass.

\begin{figure}
\includegraphics[width=8cm]{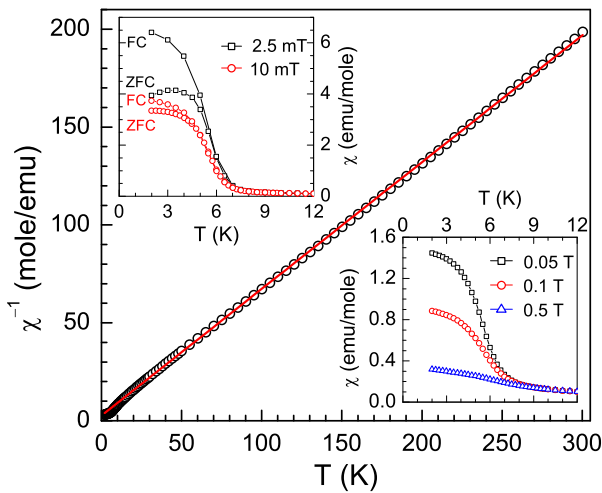}
\caption{\label{fig:fig_Chi_dc} (Colour online) The temperature dependence of dc magnetic susceptibility of PrRhSn$_{3}$, plotted as $\chi^{-1}(T)$ in the temperature range 2 -- 300 K, measured in a field of 0.5 T. The solid line represents the fit to Curie-Weiss law. The upper inset shows the zero field cooled (ZFC) and field cooled (FC) susceptibility data measured at 2.5 and 10 mT, and the lower inset shows the low temperature ZFC susceptibility data measured at 0.05, 0.1 and 0.5 T.}
\end{figure}

Figure~\ref{fig:fig_Chi_dc} shows the temperature dependence of dc magnetic susceptibility $\chi(T)$ measured under different applied magnetic fields. At low fields (e.g., at 2.5 mT and 10 mT) the magnetic susceptibility exhibits a sharp increase below 6~K, and an irreversibility between zero field cooled (ZFC) and field cooled (FC) susceptibility data below 5.5 K (upper inset of Fig.~\ref{fig:fig_Chi_dc}) which are typical of a weak ferromagnetic system and suggest a ferromagnetic cluster glass behavior consistent with the ac susceptibility observations. An increase in magnetic field strength tends to smoothen the dc magnetic susceptibility anomaly and reduces the magnitude of susceptibility (lower inset of Fig.~\ref{fig:fig_Chi_dc}). The high temperature magnetic susceptibility follows the Curie-Weiss behavior, $\chi(T) = C/(T-\theta_{\rm p})$. A linear fit to the inverse susceptibility $\chi^{-1}(T)$ vs. $T$ plot in the temperature range of 30 -- 300~K gives $C = 1.55(1)$ emu\,K/mole and $\theta_{\rm p} = -4.1(1)$~K\@. The effective moment calculated from $C$ comes out to be $\mu_{\rm eff} = 3.52(1)\,\mu_{\rm B}$, which is very close to the theoretical value of effective moment of Pr$^{3+}$ ions ($3.58\, \mu_{\rm B}$).

\begin{figure}
\includegraphics[width=8cm]{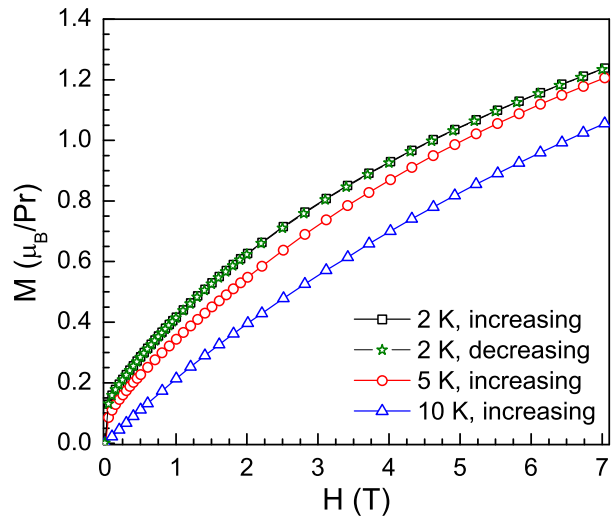}
\caption{\label{fig:fig_MH} (Colour online) The dc isothermal magnetization, $M(H)$, as a function of magnetic field measured at selected temperatures of 2, 5 and 10  K.}
\end{figure}

Figure~\ref{fig:fig_MH} shows the dc isothermal magnetization, $M(H)$ of PrRhSn$_{3}$ as a function of magnetic field measured at 2, 5 and 10~K. At 2 and 5~K, the $M(H)$ curves exhibit a small jump at very low field, revealing the appearance of spontaneous magnetization due to the formation of ferromagnetic clusters. However, no hysteresis is observed in $M(H)$ data at 2~K  recorded during increasing and decreasing cycles of magnetic field. In addition, at 2 and 5~K, the magnetization initially increases almost linearly with increasing field up to 1~T and exhibits nonlinear behavior above 1~T.  The magnetization keeps on increasing up to the investigated field of 7~T and attains only $1.25\, \mu_{\rm B}$ at 7~T which is very low compared to the theoretical saturation value of $3.2 \, \mu_{\rm B}$. The $M(H)$ curve at 10~K is slightly nonlinear, likely due to the presence of crystal field effect.

A very similar frequency dependent ac susceptibility anomaly, rapid increase in dc susceptibility below $T_{f}$ with an irreversibility in FC and ZFC susceptibility as well as a small jump in magnetization data at very small field have also been observed in the ferromagnetic cluster glass system U$_{2}$IrSi$_{3}$.\cite{Li} Cluster glass behavior has also been observed in CeNi$_{1-x}$Cu$_{x}$ ($x$ = 0.1 -- 0.7) systems which is followed by a ferromagnetic ordering due to the cluster percolative process. \cite{Solde, Marcano, MarcanoPRL}

\begin{figure}
\includegraphics[width=8cm]{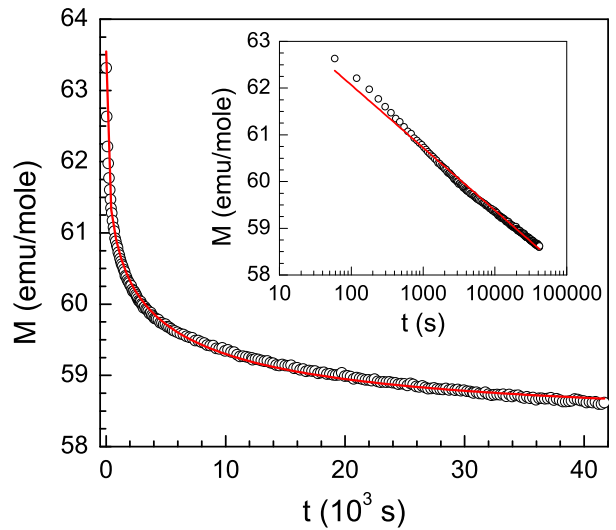}
\caption{\label{fig:fig_TRM} (Colour online) The time dependence of thermo-remnant magnetization (TRM), $M(t)$, of PrRhSn$_{3}$ measured at 2 K after switching off the cooling magnetic field of 0.05 T. The solid curve is the fit to the superposition of a stretched exponential and a constant term. The inset shows the TRM data plotted on a semi-logarithmic scale, the solid line represents the fit to logarithmic relaxation.}
\end{figure}

The thermo-remnant magnetization (TRM), $M(t)$ as a function of time is shown in Fig.~\ref{fig:fig_TRM}. To record the time dependence of TRM, the sample was field cooled in a magnetic field of 0.05~T from 50~K to 2~K and then the magnetic field was switched off and the field cooled isothermal remnant magnetization was measured at 2~K. The TRM data indicate that the magnetization decays slowly with time and remains nonzero even after 42000~s\@. The TRM data have been analyzed by logarithmic relaxation decay as well as by stretched exponential decay. The inset shows the semi-logarithmic plot of TRM data together with the fit to logarithmic decay, $M(t) = M_{0}- S \log(t)$ with $M_{0} = 64.74(3)$~emu/mole and $S = 1.34(1)$~emu/mole. The disagreement between the fit and observed TRM data suggests that these data do not follow the logarithmic relaxation. We therefore fit the observed TRM data by a superposition of a stretched exponential and a constant term, \cite{Sinha,Taniyama}
\begin{equation}
M = M_{0} + M_{1} exp\left[-\left(\frac{t}{\tau}\right)\right]^{1-n}
\label{eq:TRM}
\end{equation}
that shows a reasonable agreement between the observed data and fit. The constant term $M_{0}$ represents the longitudinal spontaneous magnetization coexisting with the frozen transverse spin component. \cite{Gabay} The solid curve in Fig.~\ref{fig:fig_TRM} represents the fit to the stretched exponential behavior (Eq.~\ref{eq:TRM}) with fitting parameters $M_{0} = 58.38(3)$~emu/mole, $M_{1} = 5.17(7)$~emu/mole, mean relaxation time $\tau = 2149(70)$~s and $n = 0.65(1)$. The above value of $M_{0}$ is equivalent to $\sim 0.01~\mu_{\rm B}$/Pr. The values of $\tau$ and $n$ are typical for the spin-glass system. The large value of $M_{0}$, again, suggests the presence of ferromagnetic clusters.

\begin{figure}
\includegraphics[width=8cm]{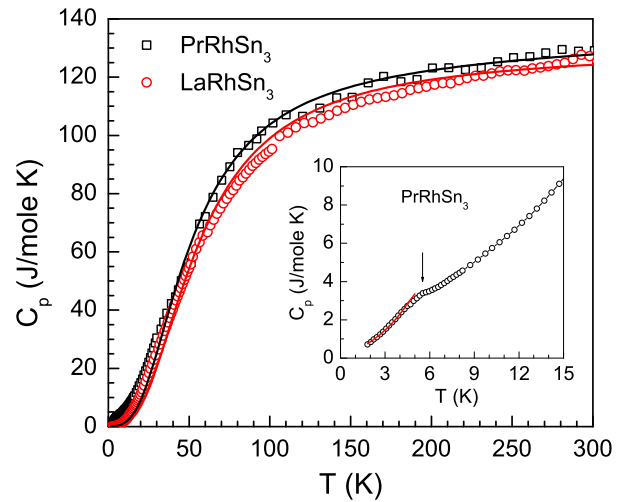}
\caption{\label{fig:fig_HC} (Colour online) The temperature dependence of specific heat $C_{\rm p}(T)$ of PrRhSn$_{3}$ and LaRhSn$_{3}$ measured in the temperature range 2 -- 300 K. The solid curves represent the fit to Debye model of lattice heat capacity. The inset shows the expanded view of low-temperature specific heat of PrRhSn$_{3}$ below 15 K. The solid curve represents the fit to $C_{\rm p}(T) = \gamma T + \beta T^{3} + \delta T^{3/2}$.}
\end{figure}

\begin{figure}
\includegraphics[width=8cm]{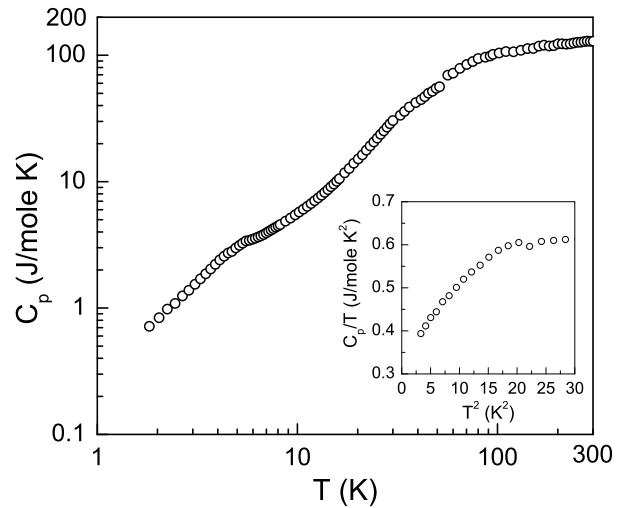}
\caption{\label{fig:fig_HC log} (Colour online) The temperature dependence of specific heat $C_{\rm p}(T)$ of PrRhSn$_{3}$ plotted on a log-log scale. The inset shows the $C_{\rm p}/T$ {\it vs}. $T^2$ plot below 5.5 K.}
\end{figure}

Figure~\ref{fig:fig_HC} shows the temperature dependence of the specific heat $C_{\rm p}(T)$ data of PrRhSn$_{3}$ and LaRhSn$_{3}$ measured at constant pressure. In order to display both the low-temperature and the high-temperature behaviors of specific heat data as well as the effect of crystal electric field more clearly we have also plotted the specific heat data of PrRhSn$_{3}$ on a log-log scale in Fig.~\ref{fig:fig_HC log}. We observe only a very weak anomaly near 5.5 K in the specific heat data of PrRhSn$_{3}$ (inset of Fig.~\ref{fig:fig_HC}) which may suggest that the magnetic susceptibility anomaly is not related to a long range magnetic order. The low-temperature specific heat data do not follow the $C_{\rm p}(T) = \gamma T + \beta T^{3}$ temperature dependence as can be seen from the $C_{\rm p}/T$ {\it vs}. $T^2$ plot (inset of Fig.~\ref{fig:fig_HC log}). However, the specific heat data below 4.5 K could be well fitted by including a magnetic term $\delta T^{3/2}$ in $C_{\rm p}(T) = \gamma T + \beta T^{3}$, i.e., by $C_{\rm p}(T) = \gamma T + \beta T^{3} + \delta T^{3/2}$ with fitting parameters $\gamma = 286(7)$~mJ/mole\,K$^{2}$, $\beta = 12.6(5)$~mJ/mole\,K$^{4}$ and $\delta  = 88.6(7)$~mJ/mole\,K$^{5/2}$, the fit is shown by the solid curve in the inset of Fig.~\ref{fig:fig_HC}. A $T^{3/2}$ temperature dependence in specific heat is typical for both ferromagnetic and spin-glass systems.\cite{Gopal,Coey,Thomson} The Sommerfeld coefficient, $\gamma = 286(7)$~mJ/mole\,K$^{2}$ obtained from the fit is very large. The large value of $\gamma$ can be attributed to the cluster-glass behaviour and/or excitonic mass enhancement by the low lying crystal field levels.

Further, we see that the specific heat attains a value of $\sim$ 127 J/mole\,K at room temperature (300 K), which is close to but slightly larger than the expected classical Dulong-Petit value of $C_{V}$ = $3nR$ = 15$R$ = 124.7 J/mole\,K. We have analyzed the specific heat data in the whole temperature range (1.8 -- 300 K) using the Debye model of lattice heat capacity. The $C_{\rm p}(T)$ data were fitted to
\begin{equation}
C_{\rm p}(T) = C_{\rm{V\,Debye}}(T) + \gamma T,
\label{eq:Debye_HC-fit}
\end{equation}
\noindent where $C_{\rm{V\,Debye}}$ represents the Debye lattice heat capacity due to the acoustic phonons and is given by \cite{Gopal}
\begin{equation}
C_{\rm{V\,Debye}}(T) =9 n R \left( \frac{T}{\Theta_{\rm{D}}} \right)^3 {\int_0^{\Theta_{\rm{D}}/T} \frac{x^4 e^x}{(e^x-1)^2}\, dx},
\label{eq:Debye_HC}
\end{equation}
where $\Theta_{\rm D}$ is the Debye temperature. The solid curve in Fig.~\ref{fig:fig_HC} represents the least squares fit of $C_{\rm p}(T)$ data by Eqs.~(\ref{eq:Debye_HC-fit}) and  (\ref{eq:Debye_HC}) using the Pad\'{e} approximant fitting function for $C_{\rm{V\,Debye}}(T)$. \cite{Ryan} The fitting parameters are $\gamma = 20(2)$~mJ/mole\,K$^{2}$ and Debye temperature $\Theta_{\rm D} = 204(2)$~K. This value of $\gamma$ is different from the above value of $\gamma$ obtained from the low-$T$ specific heat data because the Debye model of lattice heat capacity best describes the high-$T$ data where the phonon contribution due to lattice vibration is the dominating contribution, and does not properly account for the low-$T$ data. Further, we observe that the experimental data and the Debye model fit significantly deviates below 40~K. This difference arises due to the CEF contribution which is not taken care by the above analysis of specific heat. A similar analysis of specific heat data of LaRhSn$_{3}$ within the Debye model of lattice heat capacity gives $\gamma = 5(1)$~mJ/mole\,K$^{2}$ and Debye temperature $\Theta_{\rm D} = 224(1)$~K.

\begin{figure}
\includegraphics[width=8.5cm]{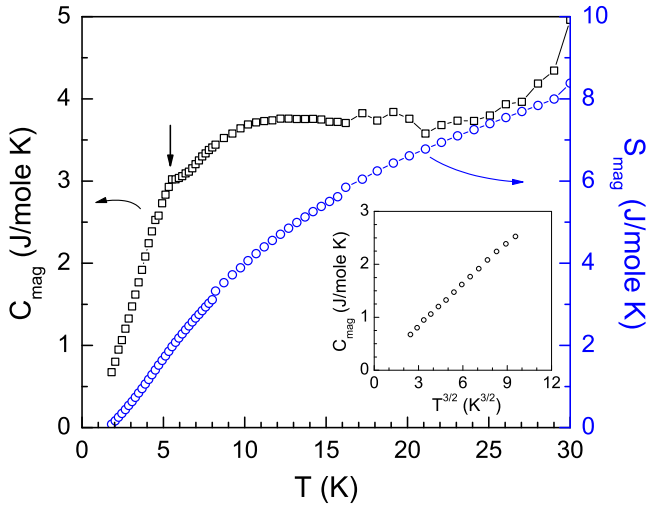}
\caption{\label{fig:fig_C_mag} (Colour online) The temperature dependence of magnetic contributions to the specific heat $C_{\rm mag}(T)$ and entropy $S_{\rm mag}(T)$ of PrRhSn$_{3}$. The inset shows the $C_{\rm mag}$ {\it vs}. $T^{3/2}$ plot below 4.5~K.}
\end{figure}

Figure~\ref{fig:fig_C_mag} shows the magnetic contribution to the specific heat of PrRhSn$_{3}$, $C_{\rm mag}$, which was obtained by subtracting the lattice contribution from the specific heat of PrRhSn$_{3}$ which we took equal to that of LaRhSn$_{3}$. The magnetic contribution to the entropy, $S_{\rm mag}$, which was obtained by integrating the $C_{\rm mag}/T$ \textit{vs}. $T$ plot is also shown in Fig.~\ref{fig:fig_C_mag}.  A weak anomaly near 5.5~K is evident in the magnetic part of the specific heat. A plot of $C_{\rm mag}$ {\it vs}. $T^{3/2}$ exhibits linear behavior (inset of Fig.~\ref{fig:fig_C_mag}) below 4.5~K revealing the $T^{3/2}$ temperature dependence of $C_{mag}$. The temperature dependence of the magnetic entropy reveals that at 5.5~K the magnetic entropy is very small which seems to be too small for the development of the long range ferromagnetic order. However, such a weak anomaly in specific heat that contains very small magnetic entropy can originate from the formation of ferromagnetic cluster state in PrRhSn$_{3}$. A similar type of weak specific heat anomaly near $T_{f}$ has also been observed in U$_{2}$IrSi$_{3}$ that is attributed to the formation of ferromagnetic cluster state. \cite{Li} Further, we also observe a broad Schottky type anomaly in magnetic part of specific heat which we attribute to the crystal electric field (CEF) effect. The very small value of magnetic entropy below $T_{f}$ suggests a CEF-split singlet ground state in PrRhSn$_{3}$. Further, the magnetic entropy attains a value of  $R \ln3$ near 38~K, which considering that in a cubic environment the nine fold degenerate multiplet ($J = 4$) of $Pr^{3+}$ splits into a combination of one singlet, one doublet and two triplets, would imply the possibility of the first excited state being a doublet around 30~K.

The magnetic entropy of each cluster is given by,\cite{Cyrot, Fisher}
\begin{equation}
    S_{\rm cl} = k_{\rm B} \ln 2.
\label{S-cluster}
\end{equation}
 If there are $N_{\rm cl}$ clusters then the total entropy associated with the clusters is $S_{\rm mag} \approx N_{\rm cl}\,S_{\rm cl}$. If we assume that on an average each cluster consists of $N_{\rm s}$ number of spins then total number of spins is $N_{\rm spin} = N_{\rm cl}\,N_{\rm s}$\@. Thus the magnetic entropy per spin \cite{Miranda}
\begin{equation}
   \frac{S_{\rm mag}}{N_{\rm spin}} \approx \frac {N_{\rm cl}\,S_{\rm cl}}{N_{\rm cl}\,N_{\rm s}} \approx \frac{ S_{\rm cl}}{N_{\rm s}} \sim \frac{k_{\rm B} \ln 2}{N_{\rm s}}.
\label{S-spin}
\end{equation}
Hence the entropy per mole of spins will be
\begin{equation}
    \frac{S_{\rm mag}}{N_{\rm spin}({\rm mol})} \approx \frac{k_{\rm B} N_{\rm A} \ln 2}{N_{\rm s}} = \frac{R \ln 2}{N_{\rm s}}.
\label{Mole-S}
\end{equation}
From the temperature dependence of the magnetic entropy we obtain $S_{\rm mag} = 1.9(1)$~J/mole\,K at $T = 5.5$~K which is $\approx  1/3$ of $R \ln 2$, thus from Eq.~(\ref{Mole-S}), $N_{\rm s} \approx   3$.  This gives an estimate of typical cluster size of about 3 spins. However, for a better estimate of cluster size small-angle neutron scattering measurement is required.

Furthermore, we have estimated the Sommerfeld-Wilson ratio,
\begin{equation}
   R_{\rm W} = \frac{\chi_0/\mu_{\rm eff}^2}{\gamma_0/\pi^2 k_{\rm B}^2},
\label{R-W}
\end{equation}
which using the value of $\chi_0 \approx 4$~emu/mole from the zero field cooled dc magnetic susceptibility measured with 2.5~mT field, $\mu_{eff} = 3.52(1)\,\mu_{B}$, and $\gamma_0=286(7)$~mJ/mole\,K$^{2}$  yields $R_{\rm W}\approx 247$\@. For a free electron gas system $R_{\rm W}=1$\@. Thus the Sommerfeld-Wilson ratio which can give an estimate of the cluster moment \cite{Miranda} at low temperature is significantly enhanced for our compound. An unusually high value of $R_{\rm W}^* \sim 700$ is reported for cluster spin-glass system ${\rm (Sr_{1-x}Ca_x)_3Ru_2O_7}$ for $x=0.2$, \cite{Qu} however, they have used slightly different expression  $R_{\rm W}^* = (\chi_0/3 \mu_{\rm B}^2)/(\gamma_0/\pi^2 k_{\rm B}^2)$ to estimate the Sommerfeld-Wilson ratio. If we use this expression then for our compound we obtain $R_{\rm W}^* \sim 1020$ which is extremely large. Recently $R_{\rm W} \sim$~20--30 were observed in Kondo cluster glass system ${\rm CePd_{1-x}Rh_x}$ for $0.8 \leq x\leq 0.87$.\cite{Westerkamp} The nearly ferromagnet systems are known to exhibit enhanced value of $R_{\rm W}$ due to Stoner enhancement, e.g.,  $R_{\rm W}=40$ for Ni$_3$Ga [\onlinecite{Julian}] and $R_{\rm W}^*=40$ for ${\rm Ca_{0.5}Sr_{0.5}RuO_4}$ [\onlinecite{Nakatsuji}]. An enhanced value of $R_{\rm W}\sim 30$ is observed due to ferromagnetic quantum critical fluctuations in ${\rm YbRh_2(Si_{0:95}Ge_{0:05})_2}$.\cite{Gegenwart} The observation of large value of $R_{\rm W}\approx247$ for PrRhSn$_3$ suggests that the electronic spin-spin interactions and the ferromagnetic fluctuations are significantly strong in PrRhSn$_3$.

\begin{figure}
\includegraphics[width=8cm]{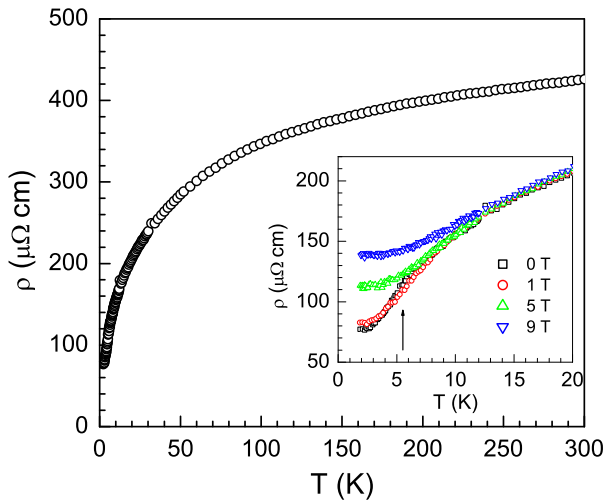}
\caption{\label{fig:fig_rho} (Colour online) The temperature dependence of electrical resistivity $\rho(T)$ of PrRhSn$_{3}$ measured in zero magnetic field. The inset shows the low temperature resistivity data measured at different fields.}
\end{figure}

\begin{figure}
\includegraphics[width=8cm]{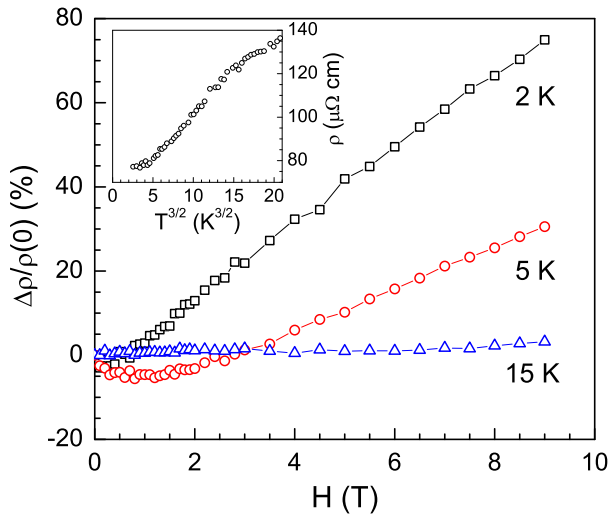}
\caption{\label{fig:fig_MR} (Colour online) The magnetic field dependence of electrical resistivity of PrRhSn$_{3}$ plotted as magnetoresistance, $\Delta \rho / \rho(0)$ at 2, 5 and 15~K. The inset shows the zero-field resistivity plotted as $\rho$ {\it vs}. $T^{3/2}$ for the temperature below 7.5~K.}
\end{figure}

Figure~\ref{fig:fig_rho} shows the temperature dependence of the electrical resistivity $\rho(T)$ data of PrRhSn$_{3}$. A metallic behavior is inferred from the temperature dependence of the electrical resistivity with a residual resistivity $\sim $ 77 $\mu \Omega$ cm (at 1.8 K) and a residual resistivity ratio of $\sim $ 6. The relatively large residual resistivity is possibly due to the presence of (micro)cracks in the sample. The electrical resistivity exhibits a broad curvature below 150 K likely due to the strong influence of crystal field effect and/or due to the band structure effect. Further, we also observe a weak anomaly near 5.5~K (inset of Fig.~\ref{fig:fig_rho}) below which the resistivity drop becomes more rapid which can be interpreted as a result of loss in spin disorder scattering due to the formation of ferromagnetic cluster glass state. The zero-field resistivity also exhibits a $T^{3/2}$ temperature dependence as can be seen from the plot of $\rho$ against $T^{3/2}$ (inset of Fig.~\ref{fig:fig_MR}) which is almost linear over the temperature range of 2.5 -- 6.0~K. It would have been better to see the temperature dependence of magnetic part of electrical resistivity data, however the non-metallic behavior of electrical resistivity of LaRhSn$_{3}$ \cite{AnandJPCM} makes it difficult to estimate the magnetic contribution to the electrical resistivity of PrRhSn$_{3}$.

We have also measured the electrical resistivity under the application of magnetic field which is shown in the inset of Fig.~\ref{fig:fig_rho}. We observe that the application of magnetic field smears out the resistivity anomaly at 5.5~K and results in an increase in the low temperature resistivity, however no effect of magnetic field is apparent above 15~K. The observation of increase in low temperature resistivity with increasing field is contrary to the behavior expected for the ferromagnetic clusters in which case one would expect a decrease in resistivity due to the alignment of magnetic moments along the field.  The magnetic field dependence of electrical resistivity plotted as magnetoresistance (MR), $\Delta\rho/\rho(0) = [\rho(H)- \rho(0)]/\rho(0)$, where $\rho(H)$ is the resistivity measured at magnetic field H, is shown in Fig.~\ref{fig:fig_MR}. The MR is negative initially up to 0.5 and 2.5~T at 2 and 5~K, respectively, and becomes positive thereafter. At 2~K and 9~T we observe a positive MR of $\sim$~75\% which is quite high. The magnitude of MR depends on temperature and is found to decrease with an increase in temperature, at 5~K and 9~T MR is $\sim$~30\% and at 15~K it is very small, only about $\sim$~2\%. The temperature 5 K is very close to the freezing temperature, therefore the MR at 5~K may be influenced by the coexisting paramagnetic phase, contributing a negative MR which may be responsible for slightly different MR behaviors at 2 and 5~K. However, the explanation for such a large positive magnetoresistance is not straight forward, we suspect that the increasing magnetic field destroys the freezing-in process and hence drives the system to the paramagnetic state thereby resulting in an increasing spin disorder scattering due to the scattering of electrons by the entirely disordered magnetic moments in paramagnetic state. \cite{Sechovsky} The formation of paramagnetic singlet state in PrRhSn$_3$ is evident in the temperature dependent susceptibility measured in applied field of 0.5T (bottom inset Fig.~\ref{fig:fig_Chi_dc}), which reveals temperature independent behaviour at low temperature as one expects for a paramagnetic singlet state.

\section{Conclusions}

We have shown that the ac and dc magnetic susceptibilities, isothermal magnetization, thermo-remnant magnetization, and specific heat data provide conclusive evidence of ferromagnetic cluster spin-glass behavior in PrRhSn$_{3}$ below a characteristic freezing temperature, $T_{f}$ = 4.3 K. The key observations include (i) a sharp anomaly in dc magnetic susceptibility with an irreversibility in ZFC and FC data, (ii) a frequency dependent cusp in ac susceptibility, (iii) a very slow decay of thermo-remnant magnetization, (iv) a small jump in isothermal magnetization, (v) a weak anomaly in specific heat with very small magnetic entropy associated with it, (vi) a $T^{3/2}$ temperature dependence in specific heat and electrical resistivity below $T_{f}$, and (vii) a very large value of Sommerfeld-Wilson ratio $R_{\rm W}\approx 247$\@. The frequency dependence of $T_{f}$ in ac susceptibility has been discussed within the framework of critical slowing down and the empirical Vogel-Fulcher law. The magnetic entropy extracted from the specific heat data reveals a CEF-split singlet ground state, and a strong influence of crystal electric field effect is observed from the specific heat and electrical resistivity data. Although at present we don't know the mechanism behind the formation of cluster-glass state in this compound, the fact that the ground state is a CEF-split singlet suggests that the underlying mechanism might have its origin in the crystal field induced frustration as in PrAu$_{2}$Si$_{2}$. Further investigations preferably by the muon spin relaxation and inelastic neutron scattering measurements are desired to understand the mechanism behind the cluster spin-glass behavior in this compound.

\section*{ACKNOWLEDGEMENTS}

We acknowledge the financial assistance from CMPC-STFC grant number CMPC-09108.


\begin{thebibliography}{48}

\bibitem{Krimmel}
 A. Krimmel, J. Hemberger, M. Nicklas, G. Knebel, W. Trinkl, M. Brando, V. Fritsch, A. Loidl, and E. Ressouche, Phys. Rev. B {\bf 59}, 6604 (1999).

\bibitem{Goremychkin}
 E. A. Goremychkin, R. Osborn, B. D. Rainford, R. T. Macaluso, D. T. Adroja, and M. Koza, Nature Physics {\bf 4}, 766 (2008).

\bibitem{Anand}
V. K. Anand, D. T. Adroja, A. D. Hillier, J. Taylor, and G. And\'{r}e, Phys. Rev. B {\bf 84}, 064440 (2011).

\bibitem{Anupam}
 Anupam, V K Anand, Z Hossain, D T Adroja, and C Geibel, J. Phys.: Condens. Matter {\bf 23}, 376001 (2011).

\bibitem{Sullow}
S. S\"{u}llow, G. J. Nieuwenhuys, A. A. Menovsky, J. A. Mydosh, S. A. M. Mentink, T. E. Mason, and W. J. L. Buyers, Phys. Rev. Lett. {\bf 78}, 354 (1997).

\bibitem{Sullow2000}
S. S\"{u}llow, S. A. M. Mentink, T. E. Mason, R. Feyerherm, G. J. Nieuwenhuys, A. A. Menovsky, and J. A. Mydosh, Phys. Rev. B {\bf 61}, 8878 (2000).

\bibitem{Mydosh}
J. A. Mydosh  \emph{Spin Glass: An Experimental Introduction} (Taylor and Francis, London, 1993).

\bibitem{Furukawa}
T. Fukuhara, I. Sakamoto, H. Sato, S. Takayanagi, and N. Wada, J. Phys.: Condens. Matter {\bf 1}, 7487 (1989).

\bibitem{Yama}
H. Yamamoto, M. Ishikawa, K. Hasegawa, and J. Sakurai, Phys. Rev. B {\bf 52}, 10136 (1995).

\bibitem{Pikul}
A. P. Pikul, D. Kaczorowski, T. Plackowski, A. Czopnik, H. Michor, E. Bauer, G. Hilscher, P. Rogl, and Yu. Grin, Phys. Rev. B {\bf 67}, 224417 (2003).

\bibitem{Naka}
M. Nakashima, K. Tabata, A. Thamizhavel, T. C. Kobayashi, M. Hedo, Y. Uwatoko, K. Shimizu, R. Settai, and Y. \={O}nuki, J. Phys.: Condens. Matter {\bf 16}, L255 (2004).

\bibitem{Pecharsky}
 V. K. Pecharsky, O. B. Hyun, and K. A. Gschneidner, Jr, Phys. Rev. B {\bf 47}, 11839 (1993).

\bibitem{Krishna}
 V. V. Krishnamurthy, K. Nagamine, I. Watanabe, K. Nishiyama, S. Ohira, M. Ishikawa, D.H. Eom, T. Ishikawa, and T.M. Briere, Phys. Rev. Lett. {\bf 88}, 046402 (2002).

\bibitem{Conti}
M. A. Continentino, S. N. de Medeiros, M. T. D. Orlando, M. B. Fontes, and E. M. Baggio- Saitovitch, Phys. Rev. B {\bf 64}, 012404 (2001).

\bibitem{Muro}
 Y. Muro, D. Eom, N. Takeda, and M. Ishikawa, J. Phys. Soc. Jpn. {\bf 67}, 3601 (1998).

\bibitem{Settai}
R. Settai, I. Sugitani, Y. Okuda, A. Thamizhavel, M. Nakashima, Y. Onuki, and H. Harima, J. Magn. Magn. Mater. {\bf 310}, 844 (2007).

\bibitem{Kimura}
N. Kimura, K. Ito, K. Saitoh, Y. Umeda, H. Aoki, and T. Terashima, Phys. Rev. Lett. {\bf 95}, 247004 (2005).

\bibitem{Kimura2007}
N. Kimura, Y. Muro, and H. Aoki, J. Phys. Soc. Jpn. {\bf 76}, 051010 (2007).

\bibitem{Sugitani}
I. Sugitani, Y. Okuda, H. Shishido, T. Yamada, A. Thamizhavel, E. Yamamoto, T. D. Matsuda, Y. Haga, T. Takeuchi, R. Settai, and Y. \={O}nuki, J. Phys. Soc. Jpn. {\bf 75}, 043703 (2006).

\bibitem{Tada}
 Y. Tada, N. Kawakami, and S. Fujimoto, Phys. Rev. B {\bf 81}, 104506 (2010).

\bibitem{AnandSSC}
 V. K. Anand, Z. Hossain and C. Geibel, Solid State Communications, {\bf 146}, 335 (2008).

\bibitem{AnandPRB}
 V. K. Anand, A. D. Hillier, D. T. Adroja, A. Strydom, H. Michor, K. A. McEwen, and B. D. Rainford, Phys. Rev. B {\bf 83}, 064522 (2011).

\bibitem{Hillier}
 A. D. Hillier, D. T. Adroja, P. Manuel, V. K. Anand, J. W. Taylor, K. A. McEwen, and B. D. Rainford, Phys. Rev. B (Submitted, 2011).

\bibitem{AnandJPCM}
 V. K. Anand, D. T. Adroja, A. D. Hillier, W. Kockelmann, A. Fraile, and A. M. Strydom, J. Phys.: Condens. Matter \textbf{23 }(2011) 276001

\bibitem{Eisenmann} B. Eisenmann, and H. Schafer, J. Less-Common Metals {\bf 123}, 89 (1986).

\bibitem{Oduchi}
Y. \={O}duchi, C. Tonohiro, A. Thamizhavel, H. Nakashima, S. Morimoto, T. D. Matsuda, Y. Haga, K. Sugiyama, T. Takeuchi, R. Settai, M. Hagiwara, and Y. \={O}nuki, J. Magn. Magn. Mater. {\bf 310}, 249 (2007).

\bibitem{Rodriguez}
\bibinfo{author}{J. Rodr\'{i}guez-Carvajal},
\bibinfo{journal} {Physica B} \textbf{\bibinfo{volume}{192}}, \bibinfo{pages} {55} (\bibinfo{year}{1993}); Program Fullprof, LLB-JRC, Laboratoire L\'{e}on Brillouin, CEA-Saclay, France,1996.

\bibitem{Schubert}
K. Schubert and, H. Pfisterer, Z. Metallkd. {\bf 41}, 433 (1950).

\bibitem{Hohenberg}
 P.C. Hohenberg, and B. I. Halperin, Rev. Mod. Phys. {\bf 49}, 435 (1977).

\bibitem{Binder}
K. Binder, and A. P. Young, Rev. Mod. Phys. {\bf 58} 801 (1986).

\bibitem{Vogel}
H. Vogel, Phys. Z. {\bf 22}, 645 (1921).

\bibitem{Fulcher}
G. S. Fulcher, J. Am. Ceram. Soc. {\bf 8}, 339 (1925).

\bibitem{Tholence}
J. L. Tholence, Solid State Commun. {\bf 35}, 113 (1980).

\bibitem{Souletie}
 J. Souletie, and J. L. Tholence, Phys. Rev. B {\bf 32}, 516(R) (1985).

\bibitem{Tholence1984}
J. L. Tholence, Physica B, {\bf 126}, 157 (1984).

\bibitem{Li}
 D. X. Li, S. Nimori, Y. Shiokawa, Y. Haga, E. Yamamoto, and Y. \={O}nuki, Phys. Rev. B {\bf 68}, 172405 (2005).

\bibitem{Solde}
 J. G. Soldevilla, J. C. G. Sal, J. A. Blanco, J. I. Espeso, and J. R. Fernandez, Phys. Rev. B \textbf{61}, 6821 (2000).

\bibitem{Marcano}
 N. Marcano, J. C. G. Sal, J. I. Espeso, L. F. Barquin, and C. Paulsen Phys. Rev. B \textbf{76}, 224419 (2007).

\bibitem{MarcanoPRL}
N. Marcano, J. C. G. Sal, J. I. Espeso, J. M. De Teresa, P. A. Algarabel, C. Paulsen, and J. R. Iglesias, Phys. Rev. Lett. \textbf{98}, 166406 (2007).

\bibitem{Sinha}
\bibinfo{author}{G. Sinha, R. Chatterjee, M. Uehara, and A. K. Majumdar},
\bibinfo{journal} {J. Magn. Magn. Mater.} \textbf{\bibinfo{volume}{164}}, \bibinfo{pages} {345} (\bibinfo{year}{1996}).

\bibitem{Taniyama} T. Taniyama, and I. Nakatani,  J. Appl. Phys. {\bf 83}, 6323 (1998).

\bibitem{Gabay} M. Gabay, and G. Toulouse, Phys. Rev. Lett. \textbf{47}, 201 (1981).

\bibitem{Gopal}
E. S. R. Gopal, \emph{Specific Heats at Low Temperatures} (Plenum, New York, 1966).

\bibitem{Coey}
J. M. D. Coey, S. von Molnar, and R. J. Gambino, Solid State Commun. {\bf 24}, 167 (1977).

\bibitem{Thomson}
J. O. Thomson, and J. R. Thomson, J. Phys. F: Metal Phys. {\bf 11}, 247 (1981).

\bibitem{Ryan}
R. J. Goetsch, V. K. Anand, A. Pandey, and D. C. Johnston, arXiv:1112.1864.

\bibitem{Cyrot}
M. Cyrot, Solid State Commun. {\bf 39}, 1009 (1981).

\bibitem{Fisher}
Daniel S. Fisher, Phys. Rev. B. \textbf{51}, 6411 (1995).

\bibitem{Miranda}
E. Miranda, and V. Dobrosavljevi\'{c}, Rep. Prog. Phys. {\bf 68}, 2337 (2005).

\bibitem{Qu}
Z. Qu, L. Spinu, H. Yuan, V. Dobrosavljevic, W. Bao, J. W. Lynn, M. Nicklas, J. Peng, T. Liu, D. Fobes, E. Flesch, and Z. Q. Mao, Phys. Rev. B \textbf{78}, 180407(R) (2008).

\bibitem{Westerkamp}
T. Westerkamp, M. Deppe, R. Küchler, M. Brando, C. Geibel, P. Gegenwart, A. P. Pikul, and F. Steglich Phys. Rev. Lett. {\bf 102}, 206404 (2009)

\bibitem{Julian}
S. R. Julian, A. P. Mackenzie, G. G. Lonzarich, C. Bergemann, R. K. W. Haselwimmer, Y. Maeno, S. Nishizaki, A. W. Tyler, S. Ikeda, and T. Fujita, Physica B {\bf 259-261}, 928 (1999).

\bibitem{Nakatsuji}
S. Nakatsuji, D. Hall, L. Balicas, Z. Fisk, K. Sugahara, M. Yoshioka, and Y. Maeno, Phys. Rev. Lett. {\bf 90}, 137202 (2003).

\bibitem{Gegenwart}
P. Gegenwart, J. Custers, Y. Tokiwa, C. Geibel, and F. Steglich, Phys. Rev. Lett. {\bf 94}, 076402 (2005).

\bibitem{Sechovsky}
V. Sechovsk\'{y}, L. Havela, K. Proke\v{s}, H. Nakotte, F. R. de Boer, E. Br\"{u}ck, J. Appl. Phys. {\bf 76}, 6913 (1994).

\end{thebibliography}
\end{document}